\documentclass[aps,prl,nofootinbib,showpacs,twocolumn]{revtex4}
\usepackage{graphicx}

\begin{document}

\title{Electron -- phonon coupling in Eliashberg -- McMillan theory
beyond adiabatic approximation}

\author{M.V. Sadovskii$^1$$^,$$^2$}

\affiliation{$^1$Institute for Electrophysics, RAS Ural Branch, 
Ekaterinburg 620016, Russia\\
$^2$M.N. Mikheev Institute for Metal Physics, RAS Ural Branch, 
Ekaterinburg 620108, Russia\\
E-mail: sadovski@iep.uran.ru}

\begin{abstract}

Eliashberg -- McMillan theory of superconductivity is essentially based on
the adiabatic approximation. Small parameter of perturbation theory is given by
$\lambda\frac{\Omega_0}{E_F}\ll 1$, where $\lambda$ is the dimensionless
electron -- phonon coupling constant, $\Omega_0$ is characteristic phonon
frequency, while $E_F$ is Fermi energy of electrons. Here we present an attempt
to describe electron -- phonon interaction within Eliashberg -- McMillan
approach in situation, when characteristic phonon frequency $\Omega_0$ becomes
large enough (comparable or exceeding the Fermi energy $E_F$). We consider the
general definition of electron -- phonon pairing coupling constant
$\lambda$, taking into account the finite value of phonon frequency. Also we
obtain the simple expression for the generalized coupling constant
$\tilde\lambda$, which determines the mass renormalization, with the account of
finite width of conduction band, and describing the smooth transition from
the adiabatic regime to the region of strong nonadiabaticity. In the case of
strong nonadiabaticity, when $\Omega_0\gg E_F$, the new small parameter appears
$\lambda\frac{E_F}{\Omega_0}\sim\lambda\frac{D}{\Omega_0}\ll 1$
($D$ is conduction band half -- width), and corrections to electronic spectrum
become irrelevant. At the same time, the temperature of superconducting
transition $T_c$ in antiadiabatic limit is still determined by Eliashberg --
McMillan coupling constant $\lambda$, while the preexponential factor in the
expression for $T_c$, conserving the form typical of weak -- coupling theory,
is determined by the bandwidth (Fermi energy). For the case of interaction with
a single optical phonon we derive the single expression for $T_c$, valid both
in adiabatic and antiadiabatic regimes and describing the continuous transition
between these two limiting cases. The results obtained are discussed in the
context of superconductivity in FeSe/STO.

\end{abstract}

\pacs{74.20.-z, 74.20.Fg, 74.20.Mn, 74.25.Kc, 74.78.Fk}

\maketitle

\newpage

\section{Introduction}

Eliashberg -- McMillan superconductivity theory is the most successful approach
to microscopic description of the properties of conventional superconductors
with electron -- phonon mechanism of Cooper pairing \cite{Scal,Izy,All}.
It basic principles can be directly generalized also for the description of
non -- phonon pairing mechanism in new high -- temperature superconductors.
Recently this theory was successfully applied to the description of record
breaking superconductivity in hydrides at high pressures \cite{Grk-Krs}.

It is widely known that Eliashberg -- Mc Millan theory is essentially based on
the applicability of adiabatic approximation and Migdal's theorem \cite{Mig},
which allows the neglect of vertex corrections in calculations of electron --
phonon coupling in typical metals. In this case the correct small parameter of
perturbation theory is $\lambda\frac{\Omega_0}{E_F}\ll 1$, where
$\lambda$ is the dimensionless Eliashberg -- McMillan electron -- phonon
coupling constant, $\Omega_0$ is characteristic phonon frequency and $E_F$ is
Fermi energy of electrons. In particular, this leads to the common opinion,
that vertex corrections in this theory can be neglected even for
$\lambda > 1$, due to the validity of inequality  $\frac{\Omega_0}{E_F}\ll 1$,
characteristic for typical metals. This is certainly correct in continuous
approximation, when we neglect the effects of lattice discreteness on electron
spectrum.

The discreteness of the lattice leads to the breaking of Migdal's theorem for
$\lambda\sim 1$ due to polaronic effects \cite{Alx,Scl}. At the same time, for
the region of $\lambda <1$ we can safely neglect these effects \cite{Scl}.
In the following we shall consider only the continuous case, limiting our
discussion to not so large values of electron -- phonon coupling $\lambda$.

Recently a number of superconductors was discovered, where the adiabatic
approximation can not be considered valid, and characteristic frequencies of
phonons are of the order or even greater than Fermi energy. We bear in mind
mainly superconductors based on FeSe monolayers, mostly the systems like
single -- atomic layer of FeSe on the SrTiO$_3$ substrate (FeSe/STO) \cite{UFN}.
For these systems this was first noted by Gor'kov \cite{Gork_1,Gork_2}, while
discussing the idea of possible $T_c$ enhancement in FeSe/STO due to interaction
with high -- energy optical phonons in SrTiO$_3$ \cite{UFN}.

\section{Self -- energy and electron -- phonon coupling constant}

\begin{figure}
\includegraphics[clip=true,width=0.5\textwidth]{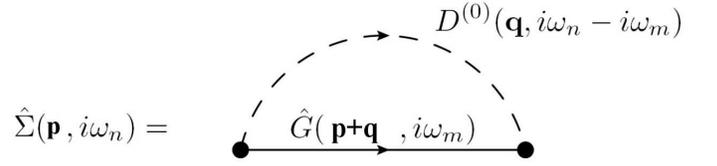}
\caption{Second -- order diagram for self -- energy. Dashed line --- phonon
Green's function $D^{(0)}$, continuous line --- electron Green's function
$G$ in Matsubara representation.}
\label{SE}
\end{figure}

Consider the second -- order (in electron -- phonon coupling) diagram shown in
Fig. \ref{SE}. At first it is sufficient to consider a metal in normal
(non superconducting) state. We can perform our analysis either in Matsubara
technique ($T\neq 0$) or in $T=0$ technique. In particular, making all
calculations in finite temperature technique, after the analytic continuation
from Matsubara to real frequencies $i\omega_n\to\varepsilon\pm i\delta$  and in
the limit of $T=0$, the contribution of diagram Fig. \ref{SE} can be written in
the standard form \cite{Scal,Schr}:
\begin{eqnarray}
\Sigma(\varepsilon,{\bf p})=\sum_{\bf p',\alpha}|g^{\alpha}_{\bf pp'}
|^2\Biggl\{\frac{f_{\bf p'}}{\varepsilon - \varepsilon_{\bf p'}
+\Omega^{\alpha}_{\bf p-p'}-i\delta}\nonumber\\
+ \frac{1-f_{\bf p'}}
{\varepsilon-\varepsilon_{\bf p'}-\Omega^{\alpha}_{\bf p-p'} + i\delta}\Biggr\}
\label{self-energy}
\end{eqnarray}
where in notations of Fig. \ref{SE} we have ${\bf k=p}$ and ${\bf p'=p+q}$.
Here $g^{\alpha}_{\bf p,p'}$ is Fr\"ohlich electron -- phonon coupling constant,
$\varepsilon_{\bf p}$ is electronic spectrum with energy zero taken at the Fermi
level, $\Omega^{\alpha}_{\bf q}$ is phonon spectrum, $f_{\bf p}$ is the Fermi
distribution (step -- like).

In particular, for the imaginary part of self -- energy at positive frequencies
we have:
\begin{equation}
Im\Sigma(\varepsilon>0,{\bf p})=-\pi\sum_{\bf p',\alpha}|g^{\alpha}_{\bf pp'}|^2
(1-f_{\bf p'})\delta(\varepsilon-\varepsilon_{\bf p'}-\Omega^{\alpha}_{\bf p-p'})
\label{im-self-energy}
\end{equation}
In these expressions index $\alpha$ enumerates the branches of phonon spectrum.
In the following we just drop it for brevity.

Eq. (\ref{self-energy}) can be identically written as:
\begin{eqnarray}
\Sigma(\varepsilon,{\bf p})=\int d\omega\sum_{\bf p'}|g_{\bf pp'}|^2
\delta(\omega-\Omega_{\bf p-p'})\times\nonumber\\
\times\Biggl\{\frac{f_{\bf p'}}
{\varepsilon - \varepsilon_{\bf p'}+\omega-i\delta}
+ \frac{1-f_{\bf p'}}
{\varepsilon - \varepsilon_{\bf p'}-\omega + i\delta}\Biggr\}
\label{self-energy_1}
\end{eqnarray}
In Eliashberg -- McMillan approach we usually get rid of explicit momentum
dependence here by averaging the matrix element of electron -- phonon
interaction over surfaces of constant energies, corresponding to
initial and final momenta ${\bf p}$ and ${\bf p'}$, which usually reduces to
the averaging over corresponding Fermi surfaces, as phonon scattering takes
place only within the narrow energy interval close to the Fermi level, with
effective width of the order of double Debye frequency $2\Omega_D$, and in
typical metals we always have $\Omega_D\ll E_F$. This is achieved by the
following replacement:
\begin{eqnarray}
|g_{\bf pp'}|^2\delta(\omega-\Omega_{\bf p-p'})\Longrightarrow\nonumber\\
\frac{1}{N(0)}
\sum_{\bf p}\frac{1}{N(0)}\sum_{\bf p'}
|g_{\bf pp'}|^2\delta(\omega-\Omega_{\bf p-p'})\delta(\varepsilon_{\bf p})
\delta(\varepsilon_{\bf p'})\nonumber\\
\equiv\frac{1}{N(0)}\alpha^2(\omega)F(\omega)
\label{Elias}
\end{eqnarray}
where in the last expression we have introduced the {\em definition} of
Eliashberg function $\alpha^2(\omega)$ and
$F(\omega)=\sum_{\bf q}\delta(\omega-\Omega_{\bf q})$ is the phonon density
of states.

\begin{figure}
\includegraphics[clip=true,width=0.3\textwidth]{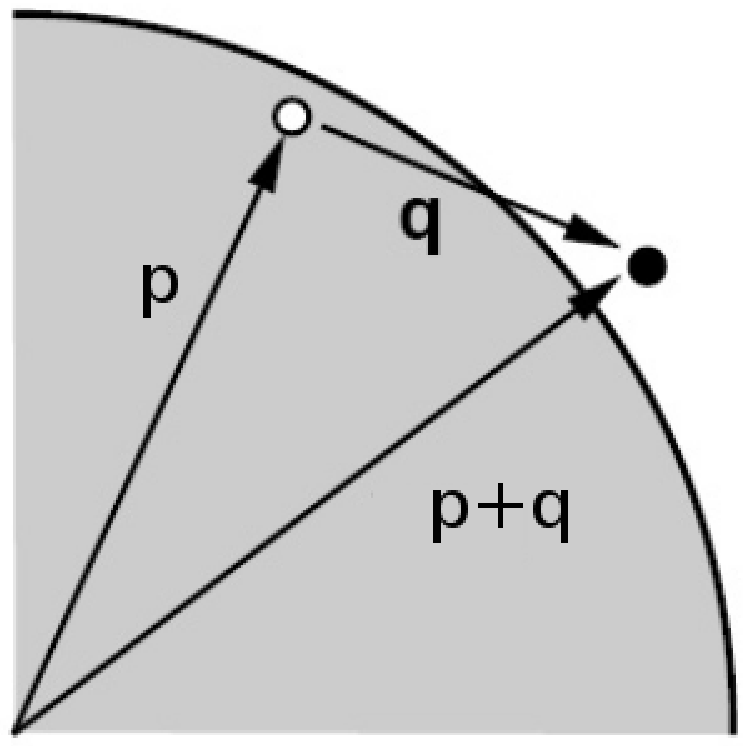}
\includegraphics[clip=true,width=0.4\textwidth]{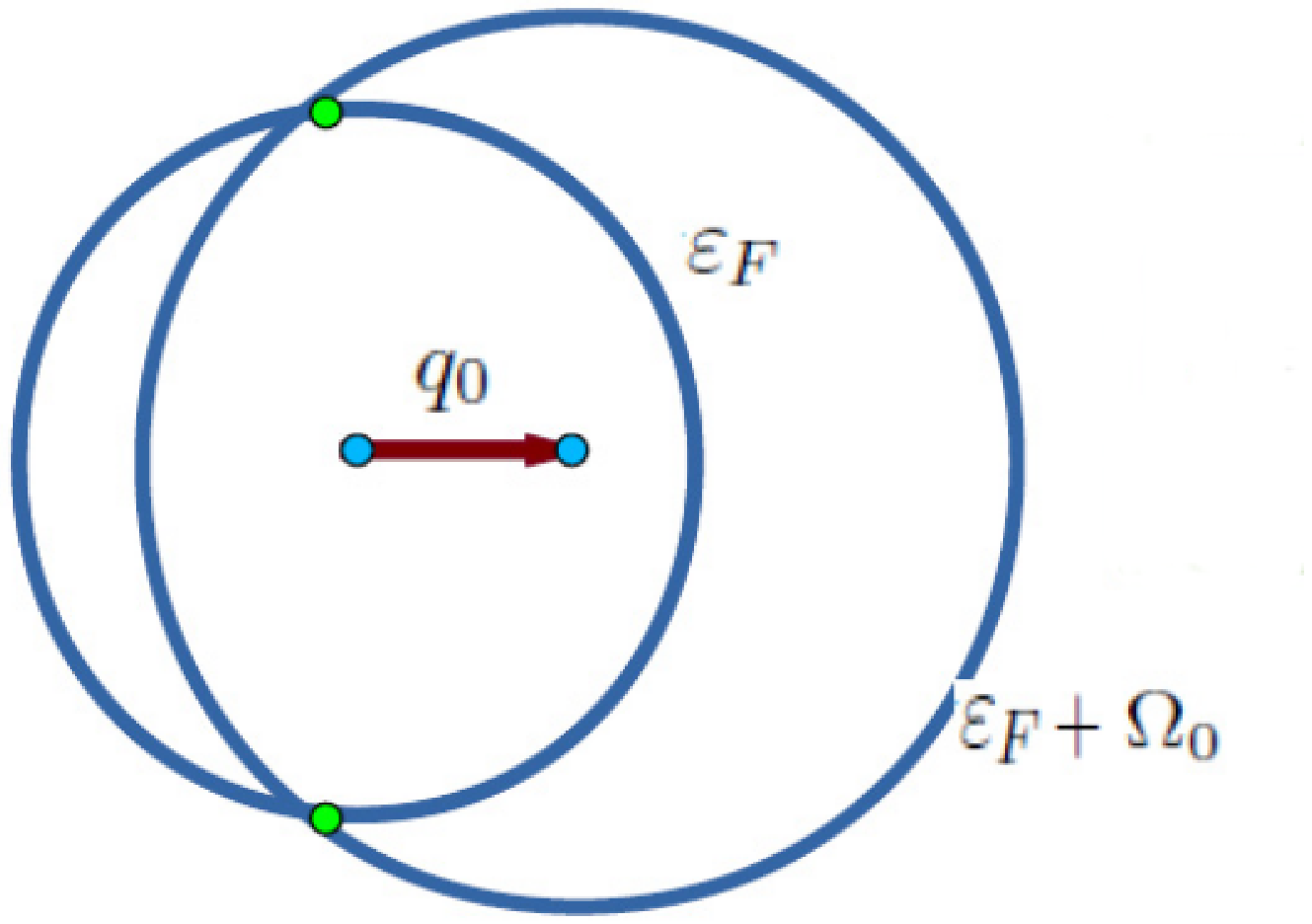}
\caption{(a) Elementary act of electron -- phonon  scattering in the vicinity
of the Fermi surface.
(b) Surfaces of constant energy for initial and final states of an electron
scattered by an optical phonon with energy comparable  to Fermi energy.
Averaging of the matrix element of interaction in (\ref{Elias_lambda}) or
(\ref{e-ph-const}) goes over the intersection region of these surfaces.}
\label{FS}
\end{figure}

In the case, when phonon energy becomes comparable with or even exceeds the
Fermi energy, electron scattering is effective not in the narrow energy layer
around the Fermi surface, but in a wider energy interval of the order of
$\Omega_0\sim E_F$, where $\Omega_0$ is a characteristic phonon frequency
(e.g. of an optical phonon). Then, for the case of initial $|{\bf p}|\sim p_F$
the averaging over ${\bf p'}$ in expression like (\ref{Elias}) should be done
over the surface of constant energy, corresponding to $E_F+\Omega_{\bf p-p'}$,
as is shown in Fig. (\ref{FS}). Then the Eq. (\ref{Elias}) is directly
generalized as:
\begin{eqnarray}
\noindent
|g_{\bf pp'}|^2\delta(\omega-\Omega_{\bf p-p'})\Longrightarrow\nonumber\\
\frac{1}{N(0)}
\sum_{\bf p}\frac{1}{N(0)}\sum_{\bf p'}
|g_{\bf pp'}|^2\times\nonumber\\
\times\delta(\omega-\Omega_{\bf p-p'})\delta(\varepsilon_{\bf p})
\delta(\varepsilon_{\bf p'}-\Omega_{\bf p-p'})\nonumber\\
\equiv\frac{1}{N(0)}\alpha^2(\omega)F(\omega)\nonumber\\
\label{Elias_1}
\end{eqnarray}
which in the last $\delta$-function simply corresponds to transition from
chemical potential $\mu$ to $\mu+\Omega_{\bf p-p'}$. We remind that, as usual,
the energy zero is taken at $\mu=0$.

After the replacement like (\ref{Elias}) or (\ref{Elias_1}) the explicit
momentum dependence of the self -- energy disappears and in fact in the
following we are dealing with Fermi surface average
$\Sigma(\varepsilon)\equiv\frac{1}{N(0)}\sum_{\bf p}
\delta(\varepsilon_{\bf p})\Sigma({\varepsilon,{\bf p}})$, which is now written
as:
\begin{eqnarray}
\Sigma(\varepsilon)=\int d\varepsilon'\int d\omega\alpha^2(\omega)F(\omega)
\Biggl\{\frac{f(\varepsilon')}
{\varepsilon - \varepsilon'+\omega-i\delta} \nonumber\\
+ \frac{1-f(\varepsilon')}
{\varepsilon - \varepsilon'-\omega + i\delta}\Biggr\}
\label{self-energy_2}
\end{eqnarray}
This expression forms the basis of Eliashberg -- McMillan theory and determines
the structure of Eliashberg equations for the description of superconductivity.

\section{Mass renormalization and electron -- phonon coupling constant}

For the case of self -- energy dependent only on frequency (and not on momentum)
we have the following simple expressions, relating mass renormalization of an
electron to the residue a the pole of the Green's function \cite{Diagr}:
\begin{equation}
Z^{-1}=1-\left.\frac{\partial\Sigma(\varepsilon)}{\partial\varepsilon}
\right|_{\varepsilon=0}
\label{Z_res}
\end{equation}
\begin{equation}
m^{\star}= \frac{m}{Z}=m\left.\left(1-\frac{\partial\Sigma(\varepsilon)}
{\partial\varepsilon}
\right|_{\varepsilon=0}\right)
\label{m_eff}
\end{equation}
Then from Eq. (\ref{self-energy_2}) by direct calculations (all integrals here
are in infinite limits) we obtain:
\begin{eqnarray}
-\left.\frac{\partial\Sigma(\varepsilon)}{\partial\varepsilon}
\right|_{\varepsilon=0}=
\int d\varepsilon'\int d\omega\alpha^2(\omega)F(\omega)
\Biggl\{\frac{f(\varepsilon')}
{(\omega-\varepsilon'-i\delta)^2}\nonumber\\
+ \frac{1-f(\varepsilon')}
{(\omega+\varepsilon'+ i\delta)^2}\Biggr\}
=2\int_{0}^{\infty}\frac{d\omega}{\omega}
\alpha^2(\omega)F(\omega)\nonumber\\
\label{derivat_0}
\end{eqnarray}
so that introducing the dimensionless Eliashberg -- McMillan electron -- phonon
coupling constant as:
\begin{equation}
\lambda=2\int_{0}^{\infty}\frac{d\omega}{\omega}\alpha^2(\omega)F(\omega)
\label{lambda_Elias_Mc}
\end{equation}
we immediately obtain the standard expression for electron mass renormalization
due to electron -- phonon interaction:
\begin{equation}
m^{\star}=m(1+\lambda)
\label{mass_ren}
\end{equation}

The function $\alpha^2(\omega)F(\omega)$ in the expression for Eliashberg --
McMillan electron -- phonon coupling constant (\ref{lambda_Elias_Mc}) should be
calculated according to (\ref{Elias}) or (\ref{Elias_1}) depending on the
relation between Fermi energy $E_F$ and characteristic phonon frequency $\Omega$
(roughly estimated by $\Omega_D$). As long as $\Omega\ll E_F$ we can use the
standard expression (\ref{Elias}), while in case of $\Omega\sim E_F$ we should
use (\ref{Elias_1}). In principle all these facts are known for a long time ---
implicitly these results were mentioned e.g. in Allen's paper \cite{Allen}, but
misunderstandings still appear \cite{Kulich}.
Using Eq. (\ref{Elias_1}) we can rewrite (\ref{lambda_Elias_Mc}) in the
following form:
\begin{eqnarray}
\lambda=\frac{2}{N(0)}\int\frac{d\omega}{\omega}
\sum_{\bf p}\sum_{\bf p'}
|g_{\bf pp'}|^2\times\nonumber\\
\times\delta(\omega-\Omega_{\bf p-p'})\delta(\varepsilon_{\bf p})
\delta(\varepsilon_{\bf p'}-\Omega_{\bf p-p'})
\label{Elias_lambda}
\end{eqnarray}
which gives the most general expression to calculate the electron -- phonon
constant $\lambda$, determining Cooper pairing in Eliashberg -- McMillan
theory.

\section{Electron interaction with optical phonons with ``forward'' scattering}

The discovery of high -- temperature superconductivity in single -- atomic
layers of FeSe on SrTiO$_3$ (FeSe/STO) and similar substrates, with record --
breaking, for iron -- based superconductors, critical temperature $T_c$,  nearly
an order of magnitude higher than in the bulk FeSe (see review in Ref.
\cite{UFN}), has sharpened the problem of search of microscopic mechanism of
$T_c$ enhancement. It was followed by the discovery in ARPES experiments on
FeSe/STO of the so  called ``replicas'' of conduction band
\cite{FeSe_ARPES_Nature}, which lead to the idea of $T_c$ enhancement due to
interaction of conduction electrons with optical phonons of SrTiO$_3$, with
rather high energies (frequencies) $\sim$ 100 meV  and ``nearly forward''
scattering (i.e. with small transferred momentum of the phonon) due to the
peculiarities of interaction with optically active Ti -- O dipoles at the
interface with STO. The model of such scattering introduced in Ref.
\cite{FeSe_ARPES_Nature} has revived the interest to earlier model of
$T_c$ enhancement, proposed by Dolgov and Kuli\'c, due to ``forward''
scattering \cite{DaDoKuO,Kulich}, which was further developed and applied to
FeSe/STO in Refs. \cite{Rade_1,Rade_2}. In fact, this model explains the
formation of the ``replicas'' of conduction band and the possibility to achieve
high values of $T_c$, though its basic conclusions were criticized (from
different points of view) in Refs. \cite{NPS_1,NPS_2,Saw} and are still under
discussion.

One of the major circumstances, which was not payed much attention in Refs.
\cite{FeSe_ARPES_Nature,Rade_1,Rade_2}, was the nonadiabatic character, as
noted by Gor'kov \cite{Gork_1,Gork_2}, of FeSe electrons interaction with
optical phonons of STO. The Fermi energy in conduction band of FeSe/STO is small,
of the order of 50-60 meV \cite{UFN,FeSe_ARPES_Nature}, which by itself is a
serious problem for theoretical explanation \cite{NPS_1,NPS_2}. Correspondingly,
the energy of optical phonons $\sim$ 100 meV exceeds is nearly twice, leading
to strong enough breaking of adiabaticity. Let us see, first of all, the
consequences of this fact for calculations of electron -- phonon coupling
constant in Eliashberg -- McMillan approach.

Consider a particular example of electrons interacting with a single optical
(Einstein -- like) phonon mode with high -- enough frequency $\Omega_0$, which
scatters essentially ``forward''. The general qualitative picture of such
scattering is shown in Fig. \ref{FS}. In this case in Eq. (\ref{Elias_lambda})
the density of phonon states is simply $F(\omega)=\delta(\omega-\Omega_0)$,
and for the momentum dependence of interaction with optical phonon at FeSe/STO
interface we can assume the characteristic dependence, obtained in Ref.
\cite{FeSe_ARPES_Nature}:
\begin{equation}
g({\bf q})=g_{0}\exp(-|{\bf q}|/q_0),
\label{g-forw}
\end{equation}
where the typical value of $q_0\sim 0.1 \frac{\pi}{a} \ll p_F$ (where $a$ is the
lattice constant and $p_F$ is the Fermi momentum), leading to nearly ``forward''
scattering of electrons by optical phonons.

Then the dimensionless pairing constant of electron -- phonon interaction in
Eliashberg theory is written as:
\begin{equation}
\lambda=\frac{2}{N(0)\Omega_0}
\sum_{\bf p}\sum_{\bf q}
|g_{\bf q}|^2\delta(\varepsilon_{\bf p})
\delta(\varepsilon_{\bf p+q}-\Omega_0)
\label{e-ph-const}
\end{equation}
As in FeSe/STO we have in fact $\Omega_0> E_F$ the finite value in the second
$\delta$-function here should be taken into account.

For simple estimates let us assume the linearized form of electronic spectrum
($ v_F $ is Fermi velocity): $\varepsilon_{\bf p}\approx v_F (| {\bf p} | -p_F)$,
which allows to perform all calculations analytically. Then, substituting
(\ref{g-forw}) into (\ref{e-ph-const}) and considering two -- dimensional case,
after calculating all integrals here we obtain \cite{NPS_2}:
\begin{equation}
\lambda=\frac{g_0^2a^2}{\pi^2v_F^2}K_1\left(\frac{2\Omega_0}{v_Fq_0}\right),
\label{lambda}
\end{equation}
where $K_1(x)$ is Bessel function of imaginary argument (McDonald function).
Using the well -- known asymptotic form of $K_1(x)$ and dropping a number of
irrelevant constants, we have:
\begin{equation}
\lambda\sim\lambda_0\frac{q_0}{4\pi p_F},
\label{lambda_0}
\end{equation}
for $\frac{\Omega_0}{v_Fq_0}\ll 1$, and
\begin{equation}
\lambda\sim\lambda_0\frac{\Omega_0}{\pi E_F}\sqrt\frac{v_Fq_0}{\Omega_0}
\exp\left(-\frac{2\Omega_0}{v_Fq_0}\right),
\label{lambda_1}
\end{equation}
for $\frac{\Omega_0}{v_Fq_0} \gg 1$.

Here we introduced the standard dimensionless electron -- phonon coupling
constant:
\begin{equation}
\lambda_0=\frac{2g_0^2}{\Omega_0}N(0),
\label{lamb_0}
\end{equation}
where $N(0)$ is the density of electronic states at the Fermi level per single
spin projection.

The result (\ref{lambda_0}) is known \cite{Rade_1,Rade_2} and by itself  is
rather unfavorable for significant $T_c$ enhancement in model under discussion.
Even worse is the situation if we take into account the large values of
$\Omega_0$, as pairing constant becomes exponentially suppressed for
$\frac{\Omega_0}{v_Fq_0}> 1$, which is typical for FeSe/STO interface, where
$\Omega_0> E_F \gg v_Fq_0 $ \cite{UFN}. This makes the enhancement of $T_c$
due to interaction of FeSe electrons with optical phonons of STO rather
improbable. In fact, similar conclusions were made from first -- principles
calculations in Ref. \cite{Johnson}, where the dependence of Eliashberg coupling
constant on frequency of the optical phonon in STO was also taken into account.
However, the effect of suppression of this constant was much smaller, which was
probably due to unrealistically large values of the Fermi energy, obtained form LDA
calculations of electronic spectrum of FeSe/STO, which does not account for the
role of correlations \cite{Johnson}. Correspondingly, in Ref. \cite{Johnson}
they always had $\Omega_0\ll E_F$. The account of correlations within
LDA+DMFT calculations, performed in Refs. \cite{NPS_1,NPS_2}, allowed to obtain
the values of Fermi energy in conduction band of FeSe/STO in accordance with
ARPES data, which show that in this system we meet with antiadiabatic situation
with $\Omega_0>E_F$.

Certainly, the results obtained above in the asymptotic of high frequencies
$\Omega_0$ depend on the form of momentum dependence in Eq. (\ref{g-forw}).
For example, if we choose the Gaussian form of interaction fall with
transferred momentum, we shall obtain more fast Gaussian suppression with
frequency in the limit of Eq. (\ref{lambda_1}). In general case, for fast enough
fall of interaction in (\ref{g-forw}) on the scale of $q_0$ we shall always
obtain rather fast suppression of coupling constant for $\Omega_0\gg v_Fq_0$.

For a more realistic case, when the optical phonon scatters electrons not
only in ``forward'' direction, but in a wide interval of transferred momenta
(as it follows e.g. from first -- principles calculations of Ref. \cite{Johnson}),
in the above expression we have simply to use large enough value of the
parameter $q_0$. Choosing e.g.  $q_0\sim 4\pi p_F$ and using the low frequency
limit (\ref{lambda_0}) we immediately obtain $\lambda\approx\lambda_0$,
i.e. the standard result. Similarly, parameter $q_0$ can be taken equal to an
inverse lattice vector $2\pi/a$. Then for $q_0\sim 2\pi/a$ from (\ref{lambda_0})
we obtain:
\begin{equation}
\lambda\sim\lambda_0\frac{1}{2p_Fa}\sim\lambda_0
\label{lambda_st}
\end{equation}
for typical $p_F\sim 1/2a$. In general case there always remain the dependence
on the value of Fermi momentum and cutoff parameter (cf. similar analysis in
Ref. \cite{Diagr}).

In the limit of (\ref{lambda_1}), assuming $q_0\sim p_F$ we immediately obtain:
\begin{equation}
\lambda\sim\frac{\sqrt{2}}{\pi}\lambda_0\sqrt\frac{\Omega_0}{E_F}
\exp\left(-\frac{\Omega_0}{E_F}\right),
\label{lambda_hifr}
\end{equation}
which simply signifies the effective interaction cutoff for $\Omega_0>E_F$ in
antiadiabatic limit. This fact was stressed by Gor'kov in Refs.
\cite{Gork_1,Gork_2}.

\section{Effects of finite bandwidth and antiadiabatic limit}

As was already noted above, the usual Migdal -- Eliashberg approach is totally
justified in adiabatic approximation, related for usual electron -- phonon
systems (metals) with the presence of a small parameter
 $\Omega_D/E_F\ll 1$ (or $\Omega_0/E_F\ll 1$ for the case of a single optical
phonon with frequency $\Omega_0$). The true parameter of perturbation theory in
this case is given by $\lambda(\Omega_0/E_F)$, which is small even for
$\lambda\sim 1$. The presence of this small parameter allows us to limit
ourselves to calculations of a simple diagram of the second order over electron
-- phonon interaction considered above, and neglect all vertex corrections
(Migdal's theorem) \cite{Mig}. These conditions are broken in FeSe/STO system,
where $\Omega_0\sim 2E_F$.

Our discussion up to now implicitly assumed the conduction band of an infinite
width. It is clear that in case of large enough characteristic phonon frequency
it may be comparable not only with Fermi energy, but also with conduction band
width. Below we shall show that in the limit of strong nonadiabaticity, when
$\Omega_0\gg E_F\sim D$ (where $D$ is the conduction band half-width), in fact,
we are dealing with the situation, when a new small parameter of perturbation
theory $\lambda D/\Omega_0\sim\lambda E_F/\Omega_0$ appears in the theory.

Let us consider the case of conduction band of the finite width $2D$ with
constant density of states (which formally corresponds formally to two --
dimensional case). The Fermi level as above is considered as an origin of
energy scale and we assume the typical case of half -- filled band.
Then (\ref{self-energy_2}) reduces to:
\begin{eqnarray}
\noindent
\Sigma(\varepsilon)=\int_{-D}^{D} d\varepsilon'\int d\omega\alpha^2(\omega)
F(\omega)\Biggl\{\frac{f(\varepsilon')}
{\varepsilon - \varepsilon'+\omega-i\delta}\nonumber\\
+ \frac{1-f(\varepsilon')}
{\varepsilon - \varepsilon'-\omega + i\delta}\Biggr\}=\nonumber\\
\noindent
=\int_{0}^{D} d\varepsilon'\int d\omega\alpha^2(\omega)F(\omega)
\Biggl\{\frac{1}
{\varepsilon + \varepsilon'+\omega-i\delta}\nonumber\\
+ \frac{1}
{\varepsilon - \varepsilon'-\omega + i\delta}\Biggr\}=\nonumber\\
=\int d\omega\alpha^2(\omega)F(\omega)
\Biggl\{\ln\frac{\varepsilon+D+\omega-i\delta}{\varepsilon-D-\omega+i\delta}
\nonumber\\
\noindent
-\ln\frac{\varepsilon+\omega-i\delta}{\varepsilon-\omega+i\delta}\Biggr\}
\label{A1}
\nonumber\\
\end{eqnarray}
For the model of a single optical phonon $F(\omega)=\delta(\omega-\Omega_0)$ and
we immediately obtain:
\begin{eqnarray}
\Sigma(\varepsilon)=\alpha^2(\Omega_0)F(\Omega_0)\Biggl\{
\ln\frac{\varepsilon+D+\Omega_0-i\delta}{\varepsilon-D-\Omega_0+i\delta}
\nonumber\\
\noindent
-\ln\frac{\varepsilon+\Omega_0-i\delta}{\varepsilon-\Omega_0+i\delta}
\biggr\}=\nonumber\\
=\alpha^2(\Omega_0)F(\Omega_0)\ln\Biggl\{\frac{\varepsilon+D+
\Omega_0-i\delta}{\varepsilon-D-\Omega_0+i\delta}
\frac{\varepsilon-\Omega_0+i\delta}{\varepsilon+\Omega_0-i\delta}\Biggr\}
\label{A2}
\end{eqnarray}
Correspondingly, form (\ref{A1}) we get:
\begin{eqnarray}
-\left.\frac{\partial\Sigma(\varepsilon)}{\partial\varepsilon}
\right|_{\varepsilon=0}
=2\int_{0}^{D}d\varepsilon'\int_{0}^{\infty}d\omega\alpha^2(\omega)F(\omega)
\frac{1}{(\omega+\varepsilon')^2}
\nonumber\\
\noindent
=2\int_{0}^{\infty}d\omega\alpha^2(\omega)F(\omega)\frac{D}{\omega(\omega+D)}
\nonumber\\
\label{A5}
\end{eqnarray}
and we can define the the generalized coupling constant as:
\begin{equation}
\tilde\lambda=2\int_{0}^{\infty}\frac{d\omega}{\omega}\alpha^2(\omega)
F(\omega)\frac{D}{\omega+D}
\label{A6}
\end{equation}
which for $D\to\infty$ reduces to the usual Eliashberg -- McMillan constant
(\ref{lambda_Elias_Mc}), while for $D\to 0$ is gives the ``antiadiabatic''
coupling constant:
\begin{equation}
\lambda_D=
2D\int \frac{d\omega}{\omega^2}\alpha^2(\omega)F(\omega)
\label{derivata_b}
\end{equation}
Eq. (\ref{A6}) describes the smooth transition between the limits of wide and
narrow conduction bands. Mass renormalization in general case is determined by
$\tilde\lambda$:
\begin{equation}
m^{\star}=m(1+\tilde\lambda)
\label{mass_renrm}
\end{equation}
In strong antiadiabatic limit of $D\ll\Omega_0$, after elementary calculations
we obtain from (\ref{A1}):
\begin{equation}
Re\Sigma(\varepsilon)=2D\int d\omega\alpha^2(\omega)F(\omega)
\frac{\varepsilon}
{\varepsilon^2-\omega^2}
\label{A3}
\end{equation}
and from (\ref{A2})
\begin{equation}
Re\Sigma(\varepsilon)=\alpha^2(\Omega_0)\frac{2D\varepsilon}{\varepsilon^2-
\Omega^2_0}=\lambda_D\frac{\Omega^2_0\varepsilon}{\varepsilon^2-\Omega^2_0}
\label{A4}
\end{equation}

For the model of a single optical phonon with frequency $\Omega_0$ we have:
\begin{equation}
\tilde\lambda=\frac{2}{\Omega_0}\alpha^2(\Omega_0)\frac{D}{\Omega_0+D}
=\lambda\frac{D}{\Omega_0+D}= \lambda_D\frac{\Omega_0}{\Omega_0+D}
\label{A7}
\end{equation}
where Eliashberg -- McMillan constant is:
\begin{equation}
\lambda=2\int_{0}^{\infty}\frac{d\omega}{\omega}\alpha^2(\omega)F(\omega)=
\alpha^2(\Omega_0)\frac{2}{\Omega_0}
\label{lambda_Elias_Mc_opt}
\end{equation}
and $\lambda_D$ reduces to:
\begin{equation}
\lambda_D=2\alpha^2(\Omega_0)\frac{D}{\Omega_0^2}=2\alpha^2(\Omega_0)
\frac{1}{\Omega_0}\frac{D}{\Omega_0}
\label{lamb_D}
\end{equation}
where in the last expression we have introduced the new small parameter
$D/\Omega_0\ll 1$, appearing in strong antiadiabatic limit. Correspondingly,
in this limit we always have:
\begin{equation}
\lambda_D=\lambda\frac{D}{\Omega_0}\sim\lambda\frac{E_F}{\Omega_0}\ll\lambda
\label{lamb_D_Mc}
\end{equation}
so that for reasonable values of $\lambda$ (even up to a strong coupling region
of $\lambda\sim 1$) ``antiadiabatic'' coupling constant remains small.
Obviously, all vertex corrections are also small in this limit, as was shown by
direct calculations in Ref. \cite{Ikeda}, which went rather unnoticed.
Thus we come to an unexpected conclusion --- in the limit of strong
nonadiabaticity the electron -- phonon coupling becomes weak!

For imaginary part of self -- energy in strong antiadiabatic limit we easily
obtain:
\begin{eqnarray}
Im\Sigma(\varepsilon>0)=-i2\pi D\varepsilon\int d\omega\alpha^2(\omega)F(\omega)
\delta(\varepsilon^2-\omega^2)=\nonumber\\
-i2\pi D\varepsilon\int d\omega\alpha^2(\omega)F(\omega)\frac{1}{2\varepsilon}
\{\delta(\varepsilon-\omega)+\delta(\varepsilon+\omega)\}-\nonumber\\
=-i\pi D\alpha^2(\varepsilon)F(\varepsilon)
\nonumber\\
\label{Im-self-energy_4}
\end{eqnarray}
which in a single phonon model reduces to:
\begin{eqnarray}
Im\Sigma(\varepsilon>0)=-i\pi D\alpha^2(\Omega_0)\delta(\varepsilon-\Omega_0)=
\nonumber\\
=-\frac{i\pi}{2}\lambda_D\Omega^2_0\delta(\varepsilon-\Omega_0)
\label{Imse}
\end{eqnarray}
From these expressions it is clear that this imaginary part is not particularly
important in this limit (being non zero only for $\varepsilon=\Omega_0$), and
equation for the real part of electronic dispersion:
\begin{equation}
\varepsilon - \varepsilon_{\bf p} - Re\Sigma(\varepsilon)=0
\label{disp}
\end{equation}
is now:
\begin{equation}
\varepsilon-\varepsilon_{\bf p}-\alpha^2(\Omega_0)\frac{2D\varepsilon}{\varepsilon^2-
\Omega^2_0}=0
\label{disp_eq}
\end{equation}
Correspondingly, for $\varepsilon\sim\varepsilon_{\bf p}$ we can write:
\begin{equation}
\varepsilon-\varepsilon_{\bf p}-\alpha^2(\Omega_0)\frac{2D\varepsilon_{\bf p}}
{\varepsilon_{\bf p}^2-\Omega^2_0}=0
\label{disp_equ}
\end{equation}
which for $\varepsilon_{\bf p}\to 0$ gives a small correction to the spectrum:
\begin{equation}
\varepsilon\approx \varepsilon_{\bf p}-\alpha^2(\Omega_0)\frac{2D}{\Omega^2_0}
\varepsilon_{\bf p}=
\varepsilon_{\bf p}-\lambda_D\varepsilon_{\bf p}=\varepsilon_{\bf p}(1-\lambda_D)
\label{disp_eq_antiad}
\end{equation}
obviously reducing to a small ($\lambda_D\ll 1$) renormalization of the effective
mass (\ref{mass_renrm}).

Physically, the weakness of electron -- phonon coupling in strong nonadiabatic
limit is pretty clear --- when ions move much faster than electrons, these have
no time to ``fit'' the rapidly changing configuration of ions and, in these
sense, only weakly react on their movement.

\section{Eliashberg equations and the temperature of superconducting transition}

All analysis above was performed for the normal state of a metal. The problem
arises, to which extent the results obtained can be generalized for the case of
a metal in superconducting state? In particular, what coupling constant
($\lambda$, $\tilde\lambda$ or $\lambda_D$) determines the temperature of
superconducting transition $T_c$ an antiadiabatic limit? Let us analyze the
situation within appropriate generalization of Eliashberg equations.

Taking into account that in antiadiabatic approximation vertex corrections are
irrelevant and neglecting the direct Coulomb interaction, Eliashberg equations
can be derived by calculating the diagram of Fig. \ref{SE}, where electronic
Green's function in superconducting state is taken in Nambu's matrix
representation. For real frequencies this is written in the following standard
form \cite{Izy}:
\begin{equation}
G(\varepsilon,{\bf p})=\frac{Z(\varepsilon)\varepsilon\tau_0+\varepsilon_{\bf p}
\tau_3+Z(\varepsilon)\Delta(\varepsilon)\tau_1}{Z^2(\varepsilon)\varepsilon^2-Z^2(\varepsilon)
\Delta^2(\varepsilon)-\varepsilon^2_{\bf p}}
\label{G_Nambu}
\end{equation}
which corresponds to the matrix of self -- energy:
\begin{equation}
\Sigma(\varepsilon,{\bf p})=[1-Z(\varepsilon)]\varepsilon\tau_0+Z(\varepsilon)
\Delta(\varepsilon)\tau_1
\label{SE_Nambu}
\end{equation}
where $\tau_i$ are standard Pauli matrices, while functions of mass
renormalization $Z(\varepsilon)$ and energy gap $\Delta(\varepsilon)$ are
determined from solution of integral Eliashberg equations, which in representation
of real frequencies are written as \cite{Izy}:
\begin{equation}
[1-Z(\varepsilon)]\varepsilon=-\int_{-D}^{D}d\varepsilon'
K(\varepsilon',\varepsilon)Re\frac{\varepsilon'}{\sqrt{\varepsilon'^2-
\Delta^2({\varepsilon'})}}sign\,\varepsilon
\label{Z_eq}
\end{equation}
\begin{equation}
Z(\varepsilon)\Delta(\varepsilon)=\int_{-D}^{D}K(\varepsilon',\varepsilon)
Re\frac{\Delta(\varepsilon')}{\sqrt{\varepsilon'^2-\Delta^2(\varepsilon')}}
sign\,\varepsilon
\label{D_eq}
\end{equation}
where integral equation kernel has the following form:
\begin{eqnarray}
K(\varepsilon',\varepsilon)=\frac{1}{2}\int_{0}^{\infty}d\omega\alpha^2(\omega)
F(\omega)\times\nonumber\\
\times\Biggl\{\frac{th\frac{\varepsilon'}{2T}+cth\frac{\omega}{2T}}
{\varepsilon'+\omega-\varepsilon-i\delta}
-\frac{th\frac{\varepsilon'}{2T}-cth\frac{\omega}{2T}}{\varepsilon'-
\omega-\varepsilon-i\delta}\Biggr\}
\label{Kernel}
\end{eqnarray}
The only difference here from the similar equations of Ref. \cite{Izy} is the
appearance of the finite integration limits, determined by the bandwidth, as well
as the absence of the contribution of direct Coulomb repulsion, which will not be
discussed here. In fact, Eqs. (\ref{Z_eq}) and (\ref{D_eq}) are the direct analog
of Eqs. (\ref{self-energy_2}) and (\ref{A1}) for normal metal and replace them
after the transition into superconducting phase.

To determine the temperature of superconducting transition it is sufficient, as
usual, to analyze the linearized Eliashberg equations, which are written as:
\begin{eqnarray}
[1-Z(\varepsilon)]\varepsilon =\int_{0}^{D}d\varepsilon'\int_{0}^{\infty}d\omega
\alpha^2(\omega)F(\omega)f(-\varepsilon')\times\nonumber\\
\times\left(\frac{1}{\varepsilon'+\varepsilon+\omega+i\delta}-
\frac{1}{\varepsilon'-\varepsilon+\omega-i\delta}\right)
\label{lin_Z}
\end{eqnarray}
\begin{eqnarray}
Z(\varepsilon)\Delta(\varepsilon)=\int_{0}^{D}\frac{d\varepsilon'}{\varepsilon'}
th\frac{\varepsilon}{2T_c}Re\Delta(\varepsilon')\times
\nonumber\\
\times\int_{0}^{\infty}d\omega
\alpha^2(\omega)F(\omega)
\times\nonumber\\
\times\left(\frac{1}{\varepsilon'+\varepsilon+\omega+i\delta}+
\frac{1}{\varepsilon'-\varepsilon+\omega-i\delta}\right)
\label{lin_D}
\end{eqnarray}
For us it is sufficient to consider in these equations the limit of
$\varepsilon\to 0$ and look for the solutions $Z(0)=Z$ and $\Delta(0)=\Delta$.
Then from (\ref{lin_Z}) we obtain:
\begin{eqnarray}
[1-Z]\varepsilon=-2\varepsilon\int_{0}^{\infty}d\omega\alpha^2(\omega)F(\omega)
\times\nonumber\\
\times\int_{0}^{D}\frac{d\varepsilon'}{(\varepsilon'+\omega)^2}=\nonumber\\
=-2\varepsilon\int_{0}^{\infty}d\omega\alpha^2(\omega)F(\omega)\frac{D}{\omega
(\omega+D)} \nonumber\\
\label{Z_sc_eq}
\end{eqnarray}
or
\begin{equation}
Z=1+\tilde\lambda
\label{Z_sc}
\end{equation}
where the constant $\tilde\lambda$ was defined above in Eq. (\ref{A6}). Thus,
precisely this effective constant determines mass renormalization both in normal
and superconducting phases. As was shown above, in the limit of strong
antiadiabaticity this renormalization is very small and determined by the
limiting value of $\lambda_D$ (\ref{lamb_D}).

Situation is different in Eq. (\ref{lin_D}). In the limit of $\varepsilon\to 0$,
using (\ref{Z_sc}) we immediately obtain from (\ref{lin_D}) the following
equation for $T_c$
\begin{equation}
1+\tilde\lambda=2\int_{0}^{\infty}d\omega\alpha^2(\omega)F(\omega)
\int_{0}^{D}\frac{d\varepsilon'}{\varepsilon'(\varepsilon'+\omega)}
th\frac{\varepsilon'}{2T_c}
\label{Tc}
\end{equation}
In antiadiabatic limit, when characteristic frequencies of phonons exceed the
width of the conduction band, we can neglect $\varepsilon'$ as compared to
$\omega$ in the denominator of the integrand in (\ref{Tc}), so that the
equation for $T_c$ is rewritten as:
\begin{eqnarray}
1+\tilde\lambda\approx 2\int_{0}^{\infty}d\omega\frac{\alpha^2(\omega)F(\omega)}
{\omega}\int_{0}^{D}\frac{d\varepsilon'}{\varepsilon'}
th\frac{\varepsilon'}{2T_c}=\nonumber\\
=\lambda\int_{0}^{D}\frac{d\varepsilon'}{\varepsilon'}
th\frac{\varepsilon'}{2T_c}
\label{Tc_ab}
\end{eqnarray}
where $\lambda$ is Eliashberg -- McMillan coupling constant as defined above in
Eq. (\ref{lambda_Elias_Mc}). From here we immediately obtain the BCS -- type
result:
\begin{equation}
T_c\sim D\exp\left(-\frac{1+\tilde\lambda}{\lambda}\right)
\label{Tc_anti}
\end{equation}
We have seen above, that in antiadiabatic limit we always have
$\tilde\lambda\to\lambda_D\ll\lambda$, so that in the exponent in (\ref{Tc_anti})
we can neglect it, so that the expression for $T_c$ is reduced simply to
to BCS {\em weak} coupling formula, with preexponential factor determined by the
half -- width of the band (Fermi energy), while the pairing coupling constant
in the exponential is determined the the general Eliashberg -- McMillan
expression (taking account the discussion above).

In the model with single optical phonon of frequency $\Omega_0$ Eq. (\ref{Tc_ab})
has the form:
\begin{equation}
1+\tilde\lambda=2\alpha^2(\Omega_0)
\int_{0}^{D}\frac{d\varepsilon'}{\varepsilon'(\varepsilon'+\Omega_0)}
th\frac{\varepsilon'}{2T_c}
\label{Tc_opt}
\end{equation}
Eq. (\ref{Tc_opt}) is easily solved (the intgeral here can be taken, as usual,
by partial integration) and we obtain:
\begin{eqnarray}
T_c\sim D\exp\left(-\frac{1+\tilde\lambda+\lambda\ln(\frac{D}{\Omega_0}+1)}
{\lambda}\right)=\nonumber\\
=\frac{D}{1+\frac{D}{\Omega_0}}\exp\left(-\frac{1+\tilde\lambda}
{\lambda}\right)
\label{Tc_opt_single}
\end{eqnarray}
where for $\lambda$ is naturally defined by Eq. (\ref{lambda_Elias_Mc_opt}).
We see, that in antiadiabatic regime, for $\frac{D}{\Omega_0}\ll 1$ this
expression reduces to (\ref{Tc_anti}), while in adiabatic limit
$\frac{D}{\Omega_0}\gg 1$ we obtain the usual expression for $T_c$ of
Eliashberg theory for the case of intermediate coupling:
\begin{equation}
T_c\sim \Omega_0\exp\left(-\frac{1+\lambda}{\lambda}\right)
\label{Tc_adi}
\end{equation}
Thus, Eq. (\ref{Tc_opt}) gives the unified expression for $T_c$, which is valid
both in adiabatic and antiadiabatic limits, smoothly interpolating between these
two limits.

Finally, we come to rather unexpected conclusions --- in the limit of strong
nonadiabaticity $T_c$ is determined by an expression like BCS weak coupling
theory, with preexponential determined not by a characteristic phonon frequency,
but by Fermi energy (the same conclusion was reached in a recent paper by
Gor'kov \cite{Gork_2}), while the pairing coupling constant conserves the
standard form of Eliashberg -- McMillan theory. The effective coupling constant
$\tilde\lambda$, tending in antiadiabatic limit to $\lambda_D$, determines the
mass renormalization, but not the temperature of superconducting transition.

\section{Conclusion}

In this work we have considered the electron -- phonon coupling in Eliashberg --
McMillan theory outside the limits of the standard adiabatic approximation.
We have obtained some simple expressions for interaction parameters of electrons
and phonons in the situation, when the characteristic frequency of phonons
$\Omega_0$ becomes large enough (comparable or even exceeding the Fermi energy
$E_F$). In particular, we have analyzed the general definition of the pairing
constant $\lambda$, taking into account the finite value of phonon frequency.
It was shown, that in a popular model with dominating ``forward'' scattering
it leads to to exponential suppression of the coupling constant for the
frequencies $\Omega_0\gg v_Fq_0$, where $q_0$ defines the characteristic size of
the region of transferred momenta, where electrons interact with phonons.
Similar situation appears also in the usual case, when $q_0$ is of the order of
inverse lattice vector, and phonon frequency exceeds the Fermi energy $E_F$.

We have obtained a simple expression for electron -- phonon coupling constant,
$\tilde\lambda$, determining the mass renormalization in Eliashberg -- McMillan
theory, taking into account the finite width of conduction band, which describes
the smooth transition from adiabatic regime to the region of strong
nonadiabaticity. It was shown, that under the conditions of strong nonadiabaticity,
when $\Omega_0\gg E_F$, a new small parameter
$\lambda\frac{E_F}{\Omega_0}\sim\lambda\frac{D}{\Omega_0}\ll 1$ ($D$ is the half
-- width of conduction band) appears in the theory, and corrections to electron
spectrum become, in fact, irrelevant, as well as all vertex corrections.
In fact, this allows us to apply the general Eliashberg equations outside the
limits of adiabatic approximation in strong antiadiabatic limit. Our results
show, that outside the limits of adiabatic approximation, in the limit of
strong nonadiabaticity, for superconductivity we have a weak coupling regime.
Mass renormalization is small and determined by effective coupling constant
$\lambda_D$, while the strength of the pairing interaction is determined by the
standard Eliashberg -- McMillan coupling constant $\lambda\gg\lambda_D$,
appropriately generalized with the account of finiteness of phonon frequency
(comparable or exceeding the Fermi energy). The cutoff of pairing interaction
in Cooper channel in antiadiabatic limit, as we have seen above (cf. also
Gor'kov's paper \cite{Gork_2}), takes place at the energies $\sim E_F$, in weak
approximation (supported by our estimates) possible vertex corrections are
irrelevant and for $T_c$ we can use the usual expression of BCS theory
(\ref{Tc_anti}), which was also stressed in Ref. \cite{Gork_2}. The small value
of $E_F$ in FeSe/STO system leads to the conclusion, that the only interaction
with antiadiabatic phonons of STO is insufficient to explain the experimentally
observed values of $T_c$, as far as we limit ourselves to weak coupling
approximation ant the value of $\lambda$ dies not exceed 0.25. In this case it
is necessary to take into account two pairing mechanisms, those responsible for
the formation of initial $T_{c0}$ in the bulk FeSe (phonons or spin fluctuations
in FeSe) and those enhancing the pairing due interaction with optical phonons of
STO. Appropriate estimates of $T_c$, performed in Ref. \cite{UFN,Gork_2} are in
reasonable agreement with experiments on FeSe/STO, with no use of the ideas on
pairing mechanisms with ``forward'' scattering. At the same time, our analysis
show, that the expression for $T_c$ like Eq. (\ref{Tc_anti}), which formally has
the form of weak coupling approximation of BCS theory, in reality ``works''
(in the limit of strong nonadiabaticity) also for large enough values of $\lambda$,
at least up to $\lambda\sim 1$, when polaronic effects become relevant.
Correspondingly, to explain the experimentally observed values of $T_c$ in
FeSe/STO it may be sufficient to deal only with electron interactions with
optical phonons of STO, as far as the values of $\lambda\sim 0.5$ can be
realized in this system. However, the realization of such large values of
coupling constant here seems rather doubtful in the light of our discussion
above (cf. also the results of first -- principles calculations of $\lambda$ in
Ref. \cite{Johnson}).

The separate question, which remained outside our discussion, is the account of
direst Coulomb repulsion. In standard Eliashberg -- McMillan theory, in adiabatic
approximation, when the frequency of phonons is orders of magnitude smaller,
than Fermi energy, this repulsion enters via Coulomb pseudopotential
$\mu^{\star}$, which is significantly suppressed by Tolmachev logarithm \cite{Izy}.
In antiadiabatic situation this mechanism of suppression does not operate, which
creates additional difficulties for realization of superconductivity.
In general, the problem of the possible role of Coulomb repulsion in
antiadiabatic regime of electron -- phonon coupling deserves serious further
studies.

This work was partially supported by RFBR grant No. 17-02-00015 and the Program
of Fundamental Research of the Presidium of the Russian Academy of Sciences
No. 12 ``Fundamental problems of high -- temperature superconductivity''.

%\newpage


\begin{thebibliography}{99}

\bibitem{Scal}D.J. Scalapino. In ``Superconductivity'', p. 449, Ed. by R.D.
Parks, Marcel Dekker, NY, 1969
\bibitem{Izy}S.V. Vonsovsky, Yu.A. Izyumov, E.Z. Kurmaev. Sverkhprovodimost'
perekhodnihh metallov ikh splavov i soedinenii. ``Nauka'', Moscow, 1977
[Superconductivity of Transition metals, Their Alloys and Compounds, Springer,
Berlin -- Heidelberg, 1982]
\bibitem{All}P.B. Allen, B. Mitrovi{\'c}. Solid State Physics, Vol. Vol. 37
(Eds. F. Seitz, D. Turnbull, H. Ehrenreich), Academic Press, NY, 1982, p. 1
\bibitem{Grk-Krs}L.P. Gor'kov, V.Z. Kresin. Rev. Mod. Phys. {\bf 90}, 011001 
(2018)
\bibitem{Mig}A.B. Migdal. Zh. Eksp. Teor. Fiz. {\bf 34}, 1438 (1958) 
[Sov. Phys. JETP {\bf 7}, 996 (1958)]
\bibitem{Alx}A.S. Alexandrov. A.B. Krebs. Usp. Fiz. Nauk {\bf 162}, 1 (1992)
[Physics
Uspekhi {\bf 35}, 345 (1992)]
\bibitem{Scl}I. Esterlis, B. Nosarzewski, E.W. Huang, D. Moritz, T.P. Devereux,
D.J. Scalapino, S.A. Kivelson. Phys. Rev. B {\bf 97}, 140501(R) (2018)
\bibitem{UFN}M.V. Sadovskii. Usp. Fiz. Nauk {\bf 178}, 1243 (2008) 
[Physics Uspekhi
{\bf 51}, 1243 (2008)]
\bibitem{Gork_1}L.P. Gor'kov. Phys. Rev. B{\bf 93}, 054517 (2016)
\bibitem{Gork_2}L.P. Gor'kov. Phys. Rev. B{\bf 93}, 060507 (2016)
\bibitem{Schr}J.R. Schrieffer. Teorija sverkprovodimosti. Fizmatlit, Moscow, 1968
[Theory of Superconductivity, WA Benjamin, NY, 1964]
\bibitem{Diagr}M.V. Sadovskii. Diagrammatika. ICS, Moscow-Izhevsk, 2010
[Diagrammatics, World Scientific, Singapore, 2006]
\bibitem{Allen}P.B. Allen. Phys. Rev. B {\bf 6}, 2577 (1972)
\bibitem{Kulich}M. Kuli{\'c}. ArXiv:1712.06222
\bibitem{FeSe_ARPES_Nature}J.J. Lee, F.T. Schmitt, R.G. Moore, S. Johnston, 
Y.T. Cui,W. Li, Z.K. Liu, M. Hashimoto, Y. Zhang, D.H. Lu, T.P. Devereaux, 
D.H. Lee, Z.X. Shen. Nature {\bf 515}, 245 (2014)
\bibitem{DaDoKuO}O.V. Danylenko, O.V.  Dolgov, M.L. Kuli\'c, V. Oudovenko.
Eur. J. Phys. B {\bf 9} 201 (1999)
\bibitem{Kulic}M.L. Kuli\'c, AIP Conference Proceedings {\bf 715} 75 (2004)
\bibitem{Rade_1}L. Rademaker, Y. Wang, T. Berlijn, S. Johnston. New J. Phys.
{\bf 18} 022001 (2016)
\bibitem{Rade_2}Y. Wang, K. Nakatsukasa, L. Rademaker, T. Berlijn, S. Johnston.
Supercond. Sci. Technol. {\bf 29}, 054009 (2016)
\bibitem{NPS_1}I.A. Nekrasov, N.S. Pavlov, V.V. Sadovskii. Pis'ma Zh. Eksp. Teor.
Fiz.  {\bf 105}, 354 (2017) [JETP Letters {\bf 105}, 370 (2017)]
\bibitem{NPS_2}I.A. Nekrasov, N.S. Pavlov, M.V. Sadovskii.
Zh. Eksp. Teor. Fiz. {\bf 153}, 590 (2018) [JETP {\bf 126}, 485 (2018)]
\bibitem{Saw}Fengmiao Li, G.A. Sawatzky. Phys. Rev. Lett. {\bf 120}, 237001 (2018)
\bibitem{Johnson}Y. Wang, A. Linscheid, T. Berlijn, S. Johnson.
Phys. Rev. B {\bf 93}, 134513 (2016)
\bibitem{Ikeda}M.A. Ikeda, A. Ogasawara, M.Sugihara. Phys. Lett. A {\bf 170},
319 (1992)


\end{thebibliography}
\end{document}